# Impedance Analysis of Modular Multilevel Converter Based on Harmonic State-Space Modeling Method

Jing Lyu, Qiang Chen, Xu Cai and Marta Molinas

*Abstract*—The small-signal impedance modeling of modular multilevel converter (MMC) is the key for analyzing resonance and stability of MMC-based ac power electronics systems. MMC is a converter system with a typical multi-frequency response due to its significant steady-state harmonic components in the arm currents, capacitor voltages, and control signals. Therefore, traditional small-signal modeling methods for 2-level voltage-source converters (VSCs) cannot be directly applied to the MMC. In this paper, the harmonic state-space (HSS) modeling approach is introduced to characterize the harmonic coupling behavior of the MMC. On this basis, the small-signal impedance models of the MMC, which take into account the internal dynamics of MMC, are developed according to the harmonic linearization principle. Furthermore, in order to reveal the impact of the internal dynamics and closed-loop control on the small-signal impedance of the MMC, three cases are considered in this paper, i.e., open-loop control, ac voltage closed-loop control, and circulating current closed-loop control. Finally, the analytical impedance models are verified by both simulation and experimental results.

## I. INTRODUCTION

Modular multilevel converter (MMC) is regarded as a promising solution for high-voltage/high-power applications, due to its advantages, such as modular design, high efficiency, low distortion of output voltage, easily scalable in terms of voltage levels, and so on [1], [2]. However, compared with conventional voltage-source converters (VSCs), e.g., 2-level VSCs, MMC has more complex internal dynamics [3], [4], such as harmonic circulating currents, capacitor voltage fluctuations, which could have a significant impact on the operation stability of MMC-based power electronics systems, especially for renewable energy integration applications [5]-[7].

The impedance-based stability analysis method is one of the preferred methods for stability analysis of power electronics systems, especially for multi-converter interconnected systems [8]-[11]. The impedance modeling of power converters is the prerequisite for applying the impedance-based analytical approach. Most of the research has so far focused on the impedance modeling of 2-level converters [9]-[14]. Sequence impedance models of 2-level grid-connected VSCs with current control and phase-locked loop (PLL) dynamics were developed in [9], [14]. *DQ* small-signal impedance of 2-level VSCs with closed-loop control was derived in [10]-[13]. However, very few authors have by far reported the impedance modeling of an MMC. The dc-side impedance of an MMC was derived for dc voltage ripple prediction in [15], but there is no mention of ac-side impedances. In the beginning of our previous work [6], [16], the ac-side impedance models of the MMC were developed, where, however, the capacitor voltage harmonics and high-order circulating current harmonics were ignored in the impedance modeling, resulting in imprecise impedance models. Later, the harmonic linearization method and harmonic state-space (HSS) method were introduced by the authors to develop the MMC impedance models in [17], [18], which are able to include all the steady-state harmonics of the state variables. In addition, a similar method to the HSS, so called multi-harmonic linearization method, was also proposed to model the MMC impedance in [19].

MMC is a converter system with a typical multi-frequency response due to its significant steady-state harmonic components in the arm currents, capacitor voltages, and control signals. Therefore, traditional small-signal modeling methods for 2-level VSCs cannot be directly applied to the MMC. In order to characterize the frequency coupling behavior of the MMC, a few authors have made efforts on the modeling of the MMC. Deore *et al*. [20] introduced dynamic phasor approach to model the MMC in frequency domain, which, however, results in a very large number of variables and equations. Dragan *et al*. [21], [22] derived the DQ model of the MMC in multiple DQ rotating coordinate frames, including zero sequence model, fundamental frequency model, and second harmonic model, which, however, involves lengthy algebra and is not suitable for system level study.

The HSS modeling method, which is able to build the multi-harmonic model of power converters, has been used to model the linear time-varying periodic (LTP) system in many fields of power systems [23-27]. Recently, the method is also applied to analyze the harmonic coupling and instability of 2-level VSCs [28], [29]. A LTP system in time domain can be transformed into a linear time invariant (LTI) system in frequency domain by the HSS modeling method, which can thus be suitable for small-signal impedance modeling.

This paper develops the small-signal impedance models of the MMC based on the HSS modeling method. First, the formulation of the HSS modeling is reviewed. Second, the HSS modeling of the MMC is presented. Third, based on the HSS model of the MMC, the small-signal impedance models are derived according to the harmonic linearization theory. Last, the validity of the derived impedance models is verified by both simulation and experimental measurements.

## II. FORMULATION OF HARMONIC STATE-SPACE MODEL

For any time-varying periodic signal $x(t)$, it can be written in the form of Fourier series as:

$$x(t) = \sum_{k \in \mathbb{Z}} X_k e^{jk\omega_1 t} \quad (1)$$

where $\omega_1 = 2\pi/T$, $T$ is the fundamental period of the signal, and $X_k$ is the Fourier coefficient that can be calculated by

$$X_k = \frac{1}{T} \int_{t_0}^{t_0+T} x(t) e^{-jk\omega_1 t} dt \quad (2)$$

The state-space equation of a LTP system can be expressed as

$$\dot{x}(t) = A(t)x(t) + B(t)u(t) \quad (3)$$

Based on the Fourier series and harmonic balance theory, the state-space equation in time domain can be transformed into the harmonic state-space equation in frequency domain, which is like

$$s\mathbf{X} = (\mathbf{A} - \mathbf{N})\mathbf{X} + \mathbf{B}\mathbf{U} \quad (4)$$

where $\mathbf{X}$, $\mathbf{U}$, $\mathbf{A}$, $\mathbf{B}$, and $\mathbf{N}$ are indicated as (5)~(9), respectively, of which the elements $X_h$, $U_h$, $A_h$, and $B_h$ are the Fourier coefficients of the $h$th harmonic of $x(t)$, $u(t)$, $A(t)$, and $B(t)$ in (3), respectively. Note that $\mathbf{A}$ and $\mathbf{B}$ are Toeplitz matrices, $\mathbf{N}$ is a diagonal matrix, and $I$ is an identity matrix.

$$\mathbf{X} = [X_{-h}, \cdots, X_{-1}, X_0, X_1, \cdots, X_h]^T \quad (5)$$

$$\mathbf{U} = [U_{-h}, \cdots, U_{-1}, U_0, U_1, \cdots, U_h]^T \quad (6)$$

$$\mathbf{A} = \begin{bmatrix} \ddots & & & & \ddots \\ & A_0 & A_{-1} & A_{-2} & \\ & A_1 & A_0 & A_{-1} & \\ & A_2 & A_1 & A_0 & \\ \ddots & & & & \ddots \end{bmatrix} \quad (7)$$

$$\mathbf{B} = \begin{bmatrix} \ddots & & & & \ddots \\ & B_0 & B_{-1} & B_{-2} & \\ & B_1 & B_0 & B_{-1} & \\ & B_2 & B_1 & B_0 & \\ \ddots & & & & \ddots \end{bmatrix} \quad (8)$$

$$\mathbf{N} = \mathrm{diag}[-jh\omega_1 \cdot I, \cdots, -j\omega_1 \cdot I, 0 \cdot I, j\omega_1 \cdot I, \cdots, jh\omega_1 \cdot I] \quad (9)$$

## III. HSS MODELING OF MMC

### A. Topology and Mathematical Model of MMC

Fig. 1 shows the circuit diagram of a three-phase MMC. Each phase-leg of the MMC consists of one upper and one lower arm connected in series between the dc terminals. Each arm consists of $N$ identical series-connected submodules (SMs), one arm inductor $L$, and an arm equivalent series resistor $R$. Each SM contains a half-bridge as a switching element and a dc storage capacitor $C_{SM}$. In high-voltage applications, $N$ may be as high as several hundreds. It should be noted that the SM may use a half-bridge or a full-bridge topology, which, however, doesn't affect the discussion here. Moreover, it needs to be pointed out that the modulation and voltage-balancing control have little effects on the impedance response due to their action on a cycle-by-cycle at the switching frequency.

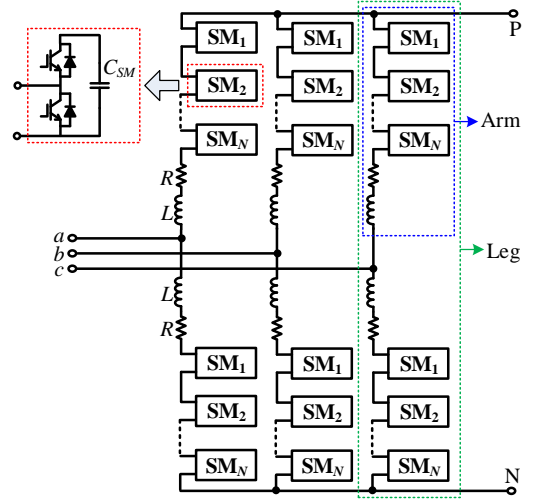

Fig. 1. Circuit diagram of a three-phase MMC.

Fig. 2 presents the average-value model of one phase leg of the MMC, where $C_{arm} = C_{SM}/N$, $C_{SM}$ is the submodule capacitance, $L$ and $R$ are the arm inductance and equivalent series resistance, $i_u$ and $i_l$ are the upper and lower arm currents, $v_{cu}^{\Sigma}$ and $v_{cl}^{\Sigma}$ are the sum capacitor voltages of the upper and lower arms, $v_g$ and $i_g$ are the ac-side phase voltage and current, $i_c$ is the circulating current, $V_{dc}$ is the dc bus voltage, and $n_u$ and $n_l$ are the insertion indices of the upper and lower arms. In addition, $Z_L$ is the equivalent load impedance on the ac-side of the MMC, and $v_p$ is the injected small perturbation voltage in order to develop the small-signal impedance models of the MMC according to the harmonic linearization theory.

Furthermore, the dc-link voltage $V_{dc}$ is assumed to be constant.

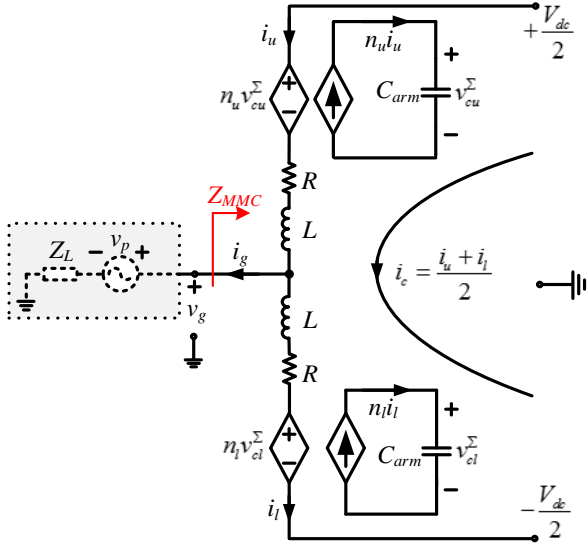

Fig. 2. Average-value model of one phase leg of MMC.

The circulating current is defined as
$$i_c = \frac{i_u + i_l}{2} \qquad (10)$$
The ac phase current can be expressed as
$$i_g = i_u - i_l \qquad (11)$$
Applying Kirchhoff's law to the single-phase equivalent circuit of the MMC in Fig. 2, one can obtain
$$v_g + L\frac{di_u}{dt} + Ri_u + v_u = \frac{V_{dc}}{2} \qquad (12)$$
$$v_g - L\frac{di_l}{dt} - Ri_l - v_l = -\frac{V_{dc}}{2} \qquad (13)$$
Based on the continuous model of the MMC [4], we have
$$\begin{cases} v_u = n_u v_{cu}^\Sigma \\ v_l = n_l v_{cl}^\Sigma \end{cases} \qquad (14)$$
$$\begin{cases} C_{arm}\dfrac{dv_{cu}^\Sigma}{dt} = n_u i_u \\ C_{arm}\dfrac{dv_{cl}^\Sigma}{dt} = n_l i_l \end{cases} \qquad (15)$$

Assuming that direct modulation is used in this work, the insertion indices of the upper and lower arms of the MMC can be expressed
$$\begin{cases} n_u = \dfrac{1}{2}\left[1 - m\cos(\omega_1 t + \theta_{m1}) - m_2\cos(2\omega_1 t + \theta_{m2})\right] \\ n_l = \dfrac{1}{2}\left[1 + m\cos(\omega_1 t + \theta_{m1}) - m_2\cos(2\omega_1 t + \theta_{m2})\right] \end{cases} \qquad (16)$$
where $m$ and $\theta_{m1}$ are the modulation index and phase of the fundamental modulation voltage, $m_2$ and $\theta_{m2}$ are the modulation index and phase of the second harmonic modulation voltage for suppressing the second harmonic circulating current. $\omega_1$ is the fundamental angular frequency.

Combining (10)-(15), the time domain state-space equation of one phase leg of the MMC can thus be expressed as:
$$\dot{x}(t) = A(t)x(t) + u(t) \qquad (17)$$
where
$$x(t) = \left[i_c, v_{cu}^\Sigma, v_{cl}^\Sigma, i_g\right]^T \qquad (18)$$
$$u(t) = \left[\frac{V_{dc}}{2L}, 0, 0, -\frac{2v_g}{L}\right]^T \qquad (19)$$
$$A(t) = \begin{bmatrix} -\dfrac{R}{L} & -\dfrac{n_u}{2L} & -\dfrac{n_l}{2L} & 0 \\ \dfrac{n_u}{C_{arm}} & 0 & 0 & \dfrac{n_u}{2C_{arm}} \\ \dfrac{n_l}{C_{arm}} & 0 & 0 & -\dfrac{n_l}{2C_{arm}} \\ 0 & -\dfrac{n_u}{L} & \dfrac{n_l}{L} & -\dfrac{R}{L} \end{bmatrix} \qquad (20)$$

B. HSS Modeling of MMC

It is noted that there are significant steady-state harmonic components in the state variables such as $i_c$, $v_{cu}^\Sigma$, and $v_{cl}^\Sigma$. In order to accurately represent the harmonic coupling effect of the MMC, the HSS modeling method is introduced to transform the state-space model of the MMC from time domain to frequency domain.

Based on the formulation of HSS model, the time domain dynamic model of (17) can be transformed into the frequency domain steady-state model of (21).
$$s\mathbf{X} = (\mathbf{A} - \mathbf{N})\mathbf{X} + \mathbf{U} \qquad (21)$$
where $\mathbf{X}$, $\mathbf{U}$, $\mathbf{A}$, and $\mathbf{N}$ are identical to (5)~(7) and (9), respectively. In addition, the elements of $\mathbf{X}$, $\mathbf{U}$, and $\mathbf{A}$ are given in (22)~(27), where the capital letters in (22)~(23) are the Fourier coefficients of the corresponding variable, while the small letters in (18)~(20) mean the time domain signals, and "$O$" is zero matrix.
$$X_{\pm h} = \left[I_{c\pm h}, V_{cu\pm h}^\Sigma, V_{cl\pm h}^\Sigma, I_{g\pm h}\right]^T \ (h \geq 0) \qquad (22)$$
$$U_0 = \left[\frac{V_{dc}}{2L}, 0, 0, 0\right]^T, U_{\pm h} = \left[0, 0, 0, -\frac{2V_{g\pm h}}{L}\right]^T \ (h \geq 1) \qquad (23)$$
$$A_0 = \begin{bmatrix} -\dfrac{R}{L} & -\dfrac{1}{4L} & -\dfrac{1}{4L} & 0 \\ \dfrac{1}{2C_{arm}} & 0 & 0 & \dfrac{1}{4C_{arm}} \\ \dfrac{1}{2C_{arm}} & 0 & 0 & -\dfrac{1}{4C_{arm}} \\ 0 & -\dfrac{1}{2L} & \dfrac{1}{2L} & -\dfrac{R}{L} \end{bmatrix} \qquad (24)$$

$$A_1 = A_{-1} = \begin{bmatrix} 0 & \dfrac{m}{8L} & -\dfrac{m}{8L} & 0 \\ -\dfrac{m}{4C_{arm}} & 0 & 0 & -\dfrac{m}{8C_{arm}} \\ \dfrac{m}{4C_{arm}} & 0 & 0 & -\dfrac{m}{8C_{arm}} \\ 0 & \dfrac{m}{4L} & \dfrac{m}{4L} & 0 \end{bmatrix} \quad (25)$$

$$A_{\pm 2} = \begin{bmatrix} 0 & \dfrac{m_2 e^{\pm j\theta_{m2}}}{8L} & \dfrac{m_2 e^{\pm j\theta_{m2}}}{8L} & 0 \\ -\dfrac{m_2 e^{\pm j\theta_{m2}}}{4C_{arm}} & 0 & 0 & -\dfrac{m_2 e^{\pm j\theta_{m2}}}{8C_{arm}} \\ -\dfrac{m_2 e^{\pm j\theta_{m2}}}{4C_{arm}} & 0 & 0 & \dfrac{m_2 e^{\pm j\theta_{m2}}}{8C_{arm}} \\ 0 & \dfrac{m_2 e^{\pm j\theta_{m2}}}{4L} & -\dfrac{m_2 e^{\pm j\theta_{m2}}}{4L} & 0 \end{bmatrix} \quad (26)$$

$$A_{3\ldots h} = A_{-3\ldots -h} = O^{4\times 4} \quad (27)$$

The steady-state solution of (21) can be calculated as

$$\mathbf{X}_{ss} = -(\mathbf{A}-\mathbf{N})^{-1}\mathbf{U} \quad (28)$$

## IV. SMALL-SIGNAL IMPEDANCE MODELING OF MMC

### A. Definition of AC-Side Small-Signal Impedance of MMC

The small-signal impedance models of the MMC are developed by means of injecting a small perturbation voltage in the ac-side of the MMC, as shown in Fig. 2. It is noted that the ac phase voltage $v_g(t)$ is perturbed by the injected small perturbation voltage $v_p(t)$ as well as the resulting perturbation current response $i_{gp}(t)$ through the load impedance $Z_L$. Therefore, the perturbation component of the ac phase voltage can be obtained

$$v_{gp}(t) = v_p(t) + Z_L i_{gp}(t) \quad (29)$$

According to the principle of harmonic linearization, the ac-side small-signal impedance of the MMC at frequency $\omega_p$ is defined as:

$$Z_{MMC}(j\omega_p) = -\dfrac{\mathbf{V}_{gp}}{\mathbf{I}_{gp}} \quad (30)$$

where $\mathbf{V}_{gp}$ and $\mathbf{I}_{gp}$ are the complex phasors of $v_{gp}(t)$ and $i_{gp}(t)$ at frequency $\omega_p$, respectively.

It is worth mentioning that, because of the frequency coupling effect of the MMC, the injected small perturbation voltage will lead to perturbations in all variables at frequencies that are listed as follows:

$$\omega_p, \omega_p \pm \omega_1, \omega_p \pm 2\omega_1, \cdots, \omega_p \pm h\omega_1 \quad (31)$$

where the harmonic order $h$ needs to be selected according to the requirements for the accuracy and complexity of the model. In general, $h=2$ is adequate for MMC to meet the accuracy requirement of the model.

To derive the small-signal impedance of the MMC, the small perturbation based HSS model of the MMC needs to be established first. Furthermore, in order to uncover the effect of the inherent dynamics and closed-loop control on the ac-side small-signal impedance of the MMC, three cases are considered in the impedance modeling in this paper, i.e., open-loop control, ac voltage closed-loop control, and circulating current closed-loop control.

### B. Impedance modeling of MMC with Open-Loop Control

If open-loop control is used, the fundamental modulation voltage is given directly and the second harmonic modulation voltage in (16) is zero. As a result, the system matrix $A(t)$ in (17) will not be influenced by the small perturbation voltage, which means that equation (17) is linear in a manner with respect to the small perturbation signals. Hence, the small perturbation based HSS model of (17) can be expressed

$$s\mathbf{X}_p = (\mathbf{A}-\mathbf{N}_p)\mathbf{X}_p + \mathbf{U}_p \quad (32)$$

where $\mathbf{X}_p$, $\mathbf{U}_p$, and $\mathbf{N}_p$ are the small perturbation components of $\mathbf{X}$, $\mathbf{U}$, and $\mathbf{N}$ in (21), respectively, which are given in (33)~(35). Additionally, $\mathbf{A}$ is given in (7), in which $A_0$ is given in (36), $A_{\pm 1}$ are identical to (25), and $A_{2\ldots h} = A_{-2\ldots -h} = O^{4\times 4}$.

$$\mathbf{X}_p = \begin{bmatrix} X_{p-h}, \cdots, X_p, \cdots, X_{p+h} \end{bmatrix}^T \quad (33)$$
$$X_{p\pm h} = \begin{bmatrix} I_{cp\pm h}, V^{\Sigma}_{cup\pm h}, V^{\Sigma}_{clp\pm h}, I_{gp\pm h} \end{bmatrix}$$

$$\mathbf{U}_p = \begin{bmatrix} U_{p-h}, \cdots, U_p, \cdots, U_{p+h} \end{bmatrix}^T \quad (34)$$
$$U_p = \begin{bmatrix} 0, 0, 0, -\dfrac{2V_p}{L} \end{bmatrix}, U_{p\pm h} = O^{1\times 4} \ (h\geq 1)$$

$$\mathbf{N}_p = \text{diag}\begin{bmatrix} j(\omega_p - h\omega_1)\cdot I^{4\times 4},\ldots, j\omega_p\cdot I^{4\times 4},\ldots, j(\omega_p + h\omega_1)\cdot I^{4\times 4} \end{bmatrix} \quad (35)$$

$$A_0 = \begin{bmatrix} -\dfrac{R}{L} & -\dfrac{1}{4L} & -\dfrac{1}{4L} & 0 \\ \dfrac{1}{2C_{arm}} & 0 & 0 & \dfrac{1}{4C_{arm}} \\ \dfrac{1}{2C_{arm}} & 0 & 0 & -\dfrac{1}{4C_{arm}} \\ 0 & -\dfrac{1}{2L} & \dfrac{1}{2L} & -\dfrac{R}{L}-\dfrac{2Z_L}{L} \end{bmatrix} \quad (36)$$

Ignoring the transient behavior of the perturbation signals, that is, letting the left-hand side of (32) to be zero, the small perturbation components of the state variables at each perturbation frequency in (32) can be solved as

$$\mathbf{X}_p = -(\mathbf{A}-\mathbf{N}_p)^{-1}\mathbf{U}_p \quad (37)$$

As a result, the ac phase current perturbation $\mathbf{I}_{gp}$ at frequency $\omega_p$ can be calculated as a function of the injected small perturbation voltage $\mathbf{V}_p$, which is expressed as

$$\mathbf{I}_{gp} = f(\mathbf{V}_p) \quad (38)$$

Additionally, the ac phase voltage perturbation $\mathbf{V}_{gp}$ at frequency $\omega_p$ can also be expressed as a function of the injected

perturbation voltage $V_p$ by substituting (38) into (29), which is expressed as

$$V_{gp} = g(V_p) \qquad (39)$$

Consequently, the ac-side small-signal impedance of the MMC with open-loop control can be obtained by substituting (38) and (39) into (30).

## C. Impedance modeling of MMC with AC Voltage Closed-Loop Control

Fig. 3 depicts the block diagram of the ac voltage closed-loop control employed in the MMC in the three-phase stationary frame, where a proportional-resonant (PR) controller is used to achieve the zero steady-state error for the sinusoidal quantities.

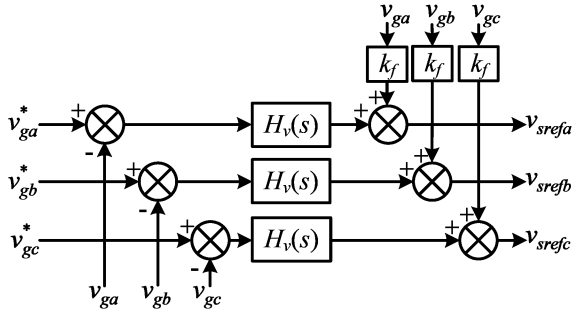

Fig. 3. Block diagram of the ac voltage closed-loop control of the MMC.

Since the ac voltage closed-loop control is employed, the fundamental modulation voltage in (16) is determined by the output of the ac voltage controller, and the second harmonic modulation voltage in (16) is zero. Therefore, equation (17) is of nonlinearity that needs to be linearized first.

The linearized time-domain state-space form of (17) is expressed

$$\dot{x}_p(t) = A(t)x_p(t) + B(t)u_p(t) \qquad (40)$$

where $x_p(t)$, $u_p(t)$, $A(t)$, and $B(t)$ are given in (41)~(44), in which the insertion indices with the subscript "0" represents the steady-state components that are given in (45), the state variables with the subscript "s" denotes the steady-state values that can be obtained by solving (28), and $G_d(s)$ represents the digital control delay of 1.5 sampling periods.

$$x_p(t) = \left[i_{cp}, v_{cup}^{\Sigma}, v_{clp}^{\Sigma}, i_{gp}\right]^T \qquad (41)$$

$$u_p(t) = \left[v_p, v_p, v_p, v_p\right]^T \qquad (42)$$

$$A(t) = \begin{bmatrix} -\dfrac{R}{L} & -\dfrac{n_{u0}}{2L} & -\dfrac{n_{l0}}{2L} & \dfrac{G_v(v_{cus}^{\Sigma} - v_{cls}^{\Sigma})}{2LV_{dc}}Z_L \\[2ex] \dfrac{n_{u0}}{C_{arm}} & 0 & 0 & \dfrac{n_{u0}}{2C_{arm}} - \dfrac{G_v\left(i_{cs} + \dfrac{i_{gs}}{2}\right)}{C_{arm}V_{dc}}Z_L \\[2ex] \dfrac{n_{l0}}{C_{arm}} & 0 & 0 & -\dfrac{n_{l0}}{2C_{arm}} + \dfrac{G_v\left(i_{cs} - \dfrac{i_{gs}}{2}\right)}{C_{arm}V_{dc}}Z_L \\[2ex] 0 & -\dfrac{n_{u0}}{L} & \dfrac{n_{l0}}{L} & -\dfrac{R}{L} + \dfrac{G_v(v_{cus}^{\Sigma} + v_{cls}^{\Sigma}) - 2V_{dc}}{LV_{dc}}Z_L \end{bmatrix} \qquad (43)$$

$$B(t) = \text{diag}\left[\dfrac{G_v(v_{cus}^{\Sigma} - v_{cls}^{\Sigma})}{2LV_{dc}}, -\dfrac{G_v\left(i_{cs} + \dfrac{i_{gs}}{2}\right)}{C_{arm}V_{dc}}, \dfrac{G_v\left(i_{cs} - \dfrac{i_{gs}}{2}\right)}{C_{arm}V_{dc}}, \dfrac{G_v(v_{cus}^{\Sigma} + v_{cls}^{\Sigma}) - 2V_{dc}}{LV_{dc}}\right] \qquad (44)$$

in which

$$\begin{cases} n_{u0} = \dfrac{1}{2}\left[1 - m\cos(\omega_1 t + \theta_{m1})\right] \\ n_{l0} = \dfrac{1}{2}\left[1 + m\cos(\omega_1 t + \theta_{m1})\right] \end{cases} \qquad (45)$$

$$G_v = \left[k_f + H_v(s)\right]G_d(s) \qquad (46)$$

Thus, the small perturbation form of the HSS model of the MMC with ac voltage closed-loop control can be expressed as

$$s\mathbf{X}_p = (\mathbf{A} - \mathbf{N}_p)\mathbf{X}_p + \mathbf{B}\mathbf{U}_p \qquad (47)$$

where $\mathbf{X}_p$ and $\mathbf{N}_p$ are identical to (32), $\mathbf{U}_p$ is given in (48). In addition, $\mathbf{A}$ and $\mathbf{B}$ are the Toeplitz matrices of the Fourier coefficients of $A(t)$ and $B(t)$ in (43) and (44), respectively, whose elements are shown in (49)~(55), in which the subscript "0", "±1", and "±2" represent the Fourier coefficient values of the state variables at frequency dc, $\pm\omega_1$, and $\pm 2\omega_1$, respectively.

$$\begin{aligned} \mathbf{U}_p &= \left[U_{p-h}, \cdots, U_p, \cdots, U_{p+h}\right]^T \\ U_p &= \left[V_p, V_p, V_p, V_p\right], U_{p\pm h} = O^{1\times 4} \ (h \geq 1) \end{aligned} \qquad (48)$$

$$A_0 = \begin{bmatrix} -\dfrac{R}{L} & -\dfrac{1}{4L} & -\dfrac{1}{4L} & \dfrac{G_v\left(V_{cu0}^{\Sigma}-V_{cl0}^{\Sigma}\right)}{2LV_{dc}}Z_L \\ \dfrac{1}{2C_{arm}} & 0 & 0 & \dfrac{1}{4C_{arm}} - \dfrac{G_v\left(I_{c0}+\dfrac{I_{g0}}{2}\right)}{C_{arm}V_{dc}}Z_L \\ \dfrac{1}{2C_{arm}} & 0 & 0 & -\dfrac{1}{4C_{arm}} + \dfrac{G_v\left(I_{c0}-\dfrac{I_{g0}}{2}\right)}{C_{arm}V_{dc}}Z_L \\ 0 & -\dfrac{1}{2L} & \dfrac{1}{2L} & -\dfrac{R}{L} + \dfrac{G_v\left(V_{cu0}^{\Sigma}+V_{cl0}^{\Sigma}\right)-2V_{dc}}{LV_{dc}}Z_L \end{bmatrix}$$

(49)

$$A_{\pm 1} = \begin{bmatrix} 0 & \dfrac{me^{\pm j\theta_{m1}}}{8L} & -\dfrac{me^{\pm j\theta_{m1}}}{8L} & \dfrac{G_v\left(V_{cu\pm 1}^{\Sigma}-V_{cl\pm 1}^{\Sigma}\right)}{2LV_{dc}}Z_L \\ -\dfrac{me^{\pm j\theta_{m1}}}{4C_{arm}} & 0 & 0 & -\dfrac{me^{\pm j\theta_{m1}}}{8C_{arm}} - \dfrac{G_v\left(I_{c\pm 1}+\dfrac{I_{g\pm 1}}{2}\right)}{C_{arm}V_{dc}}Z_L \\ \dfrac{me^{\pm j\theta_{m1}}}{4C_{arm}} & 0 & 0 & -\dfrac{me^{\pm j\theta_{m1}}}{8C_{arm}} + \dfrac{G_v\left(I_{c\pm 1}-\dfrac{I_{g\pm 1}}{2}\right)}{C_{arm}V_{dc}}Z_L \\ 0 & \dfrac{me^{\pm j\theta_{m1}}}{4L} & \dfrac{me^{\pm j\theta_{m1}}}{4L} & \dfrac{G_v\left(V_{cu\pm 1}^{\Sigma}+V_{cl\pm 1}^{\Sigma}\right)}{LV_{dc}}Z_L \end{bmatrix}$$

(50)

$$A_{\pm 2} = \begin{bmatrix} 0 & 0 & 0 & \dfrac{G_v\left(V_{cu\pm 2}^{\Sigma}-V_{cl\pm 2}^{\Sigma}\right)}{2LV_{dc}}Z_L \\ 0 & 0 & 0 & -\dfrac{G_v\left(I_{c\pm 2}+\dfrac{I_{g\pm 2}}{2}\right)}{C_{arm}V_{dc}}Z_L \\ 0 & 0 & 0 & \dfrac{G_v\left(I_{c\pm 2}-\dfrac{I_{g\pm 2}}{2}\right)}{C_{arm}V_{dc}}Z_L \\ 0 & 0 & 0 & \dfrac{G_v\left(V_{cu\pm 2}^{\Sigma}+V_{cl\pm 2}^{\Sigma}\right)}{LV_{dc}}Z_L \end{bmatrix}$$

(51)

$$A_{3\ldots h} = A_{-3\ldots -h} = O^{4\times 4} \quad (52)$$

$$B_0 = \begin{bmatrix} \dfrac{G_v\left(V_{cu0}^{\Sigma}-V_{cl0}^{\Sigma}\right)}{2LV_{dc}} & 0 & 0 & 0 \\ -\dfrac{G_v\left(I_{c0}+\dfrac{I_{g0}}{2}\right)}{C_{arm}V_{dc}} & 0 & 0 & 0 \\ \dfrac{G_v\left(I_{c0}-\dfrac{I_{g0}}{2}\right)}{C_{arm}V_{dc}} & 0 & 0 & 0 \\ \dfrac{G_v\left(V_{cu0}^{\Sigma}+V_{cl0}^{\Sigma}\right)}{LV_{dc}} - \dfrac{2}{L} & 0 & 0 & 0 \end{bmatrix}$$

(53)

$$B_{\pm h} = \begin{bmatrix} \dfrac{G_v\left(V_{cu\pm h}^{\Sigma}-V_{cl\pm h}^{\Sigma}\right)}{2LV_{dc}} & 0 & 0 & 0 \\ -\dfrac{G_v\left(I_{c\pm h}+\dfrac{I_{g\pm h}}{2}\right)}{C_{arm}V_{dc}} & 0 & 0 & 0 \\ \dfrac{G_v\left(I_{c\pm h}-\dfrac{I_{g\pm h}}{2}\right)}{C_{arm}V_{dc}} & 0 & 0 & 0 \\ \dfrac{G_v\left(V_{cu\pm h}^{\Sigma}+V_{cl\pm h}^{\Sigma}\right)}{LV_{dc}} & 0 & 0 & 0 \end{bmatrix} \quad (h=1,2) \quad (54)$$

$$B_{3\ldots h} = B_{-3\ldots -h} = O^{4\times 4} \quad (55)$$

Analogously, ignoring the transient behavior of the perturbation signals in (47), the following equation can be made by solving (47):

$$\mathbf{X}_p = -\left(\mathbf{A}-\mathbf{N}_p\right)^{-1}\mathbf{B}\mathbf{U}_p \quad (56)$$

Similar to that with open-loop control, the impedance modeling of the MMC with ac voltage closed-loop control can be implemented in the same way. According to (56) and (29), the perturbation components $V_{gp}$ and $I_{gp}$ at frequency $\omega_p$ of the ac phase voltage and current of the MMC can be expressed as a function of the injected small perturbation voltage $V_p$, respectively. Consequently, the ac-side small-signal impedance of the MMC with ac voltage closed-loop control can be obtained by substituting $V_{gp}$ and $I_{gp}$ into (30).

### D. Impedance modeling of MMC with Circulating Current Closed-Loop Control

The circulating current control strategies commonly used in the MMC can be categorized into two groups, one is based on PI controllers in the *dq* rotating frame, the other is based on PR controllers in the *abc* stationary frame. However, both of these two strategies are able to effectively suppress the harmonic circulating currents. It is worth noting that no matter the integral part of the PI controller or the resonant part of the PR controller, it can mainly affect the voltage and current signals at around the designated frequency while having little impact on those signals at other frequencies. In other words, the impedance-frequency characteristics of the MMC are mainly influenced by the proportional part of the controller. Therefore, to reveal the influence mechanism of the circulating current closed-loop control on the impedance-frequency characteristics of the MMC, a proportional controller based circulating current control [3] is used in this paper, as presented in Fig. 4.

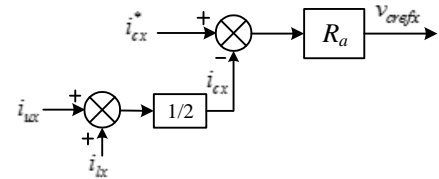

Fig. 4. Circulating current closed-loop control of the MMC.

In a similar way to that with the ac voltage closed-loop control, we can derive the ac-side small-signal impedance of the MMC with circulating current closed-loop control. Due to the limited space, the detailed derivation process is not given in this paper.

## V. Impedance Model Verification

### A. Simulation Verification

To verify the derived impedance models of the MMC, a nonlinear time domain simulation model of a three-phase MMC with a three-phase resistance load has been built by using MATLAB/Simulink. In the simulation, the ac-side small-signal impedances of the MMC are measured by means of injecting a series of small perturbation voltage signals at different frequencies in the ac-side of the MMC. Then by measuring the resulting perturbation current signals, the ac-side small-signal impedances can be easily calculated for each frequency. The main electrical parameters of the MMC in the simulation are as follows: rated power $P_N$ = 50 MW, nominal ac line voltage RMS value $V_N$ = 166 kV, nominal dc-bus voltage $V_{dc}$ = 320 kV, fundamental angular frequency $\omega_1$ = 314 rad/s, SM number per arm $N$ = 20, SM capacitance $C_{SM}$ = 140 μF, arm inductance $L$ = 360 mH, and arm resistance $R$ = 1 Ω.

Fig. 5 shows the comparison between the analytical and simulation measured ac-side small-signal impedances of the MMC with open-loop control, where the harmonic order $h$ of the analytical model is selected as 4. It is seen that the analytical impedance matches well with the measured result in the simulation, which validates the analytical impedance model. Furthermore, it can been seen that there are several resonance points in the ac-side impedance of the MMC below 200 Hz, where the resonance concave at the fundamental frequency is generated by the fundamental voltage control, and the other resonances are originated from the internal dynamics of the MMC. And, more remarkable, the resonance peak around 21 Hz caused by the internal circulating current resonance of the MMC, is likely to result in oscillations by interacting with the impedance of the load converter. The frequency at the resonance peak mainly depends on the main circuit parameters of the MMC, such as the arm inductance, submodule capacitance, submodule number per arm, etc.

Fig. 6 demonstrates the impact of the harmonic order $h$ of the HSS model on the accuracy of the analytical impedance model of the MMC. It can be seen that the higher the harmonic order $h$ of the HSS model is, the higher the accuracy of the analytical impedance model is. However, the harmonic order of the HSS model needs to be determined from the trade-off between the accuracy and complexity of the model. In general, the harmonic order $h$=4 is sufficient for the accuracy requirement.

Fig. 7 presents the comparison between the analytical and simulation measured ac-side small-signal impedances of the MMC with ac voltage closed-loop control (where $K_{pv}$=1, $K_{rv}$=20), which verifies the analytical impedance model as well. Furthermore, by comparing Fig. 7 and Fig. 5, it can be observed that the shape of the ac-side small-signal impedance with ac voltage closed-loop control is very similar to that with open-loop control, where the major difference is at the funda-mental frequency (50 Hz hereof) because of the integral part of the ac voltage controller. Besides, it can be seen from Fig. 8 how the proportional gain of the ac voltage controller affects the ac-side small-signal impedance of the MMC, where the larger the proportional gain of the ac voltage controller is, the smaller the magnitude of the MMC impedance in the entire frequency range is, but having little impact on the resonance frequencies. In addition, it needs to be pointed out that the resonant gain of the ac voltage controller has less effect on the MMC impedance except at the fundamental frequency.

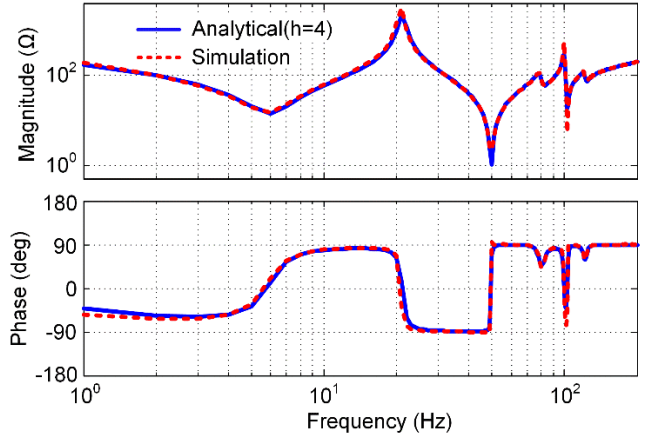

Fig. 5. Comparison between the analytical and measured ac-side impedances of the MMC with open-loop control.

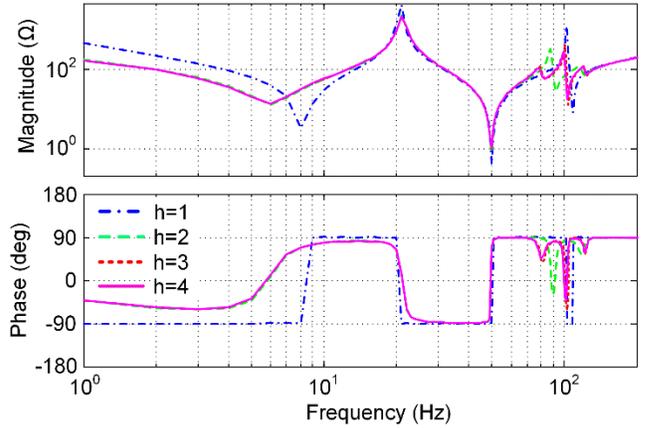

Fig. 6. Impact of the harmonic order of the HSS model on the accuracy of the analytical impedance model of the MMC.

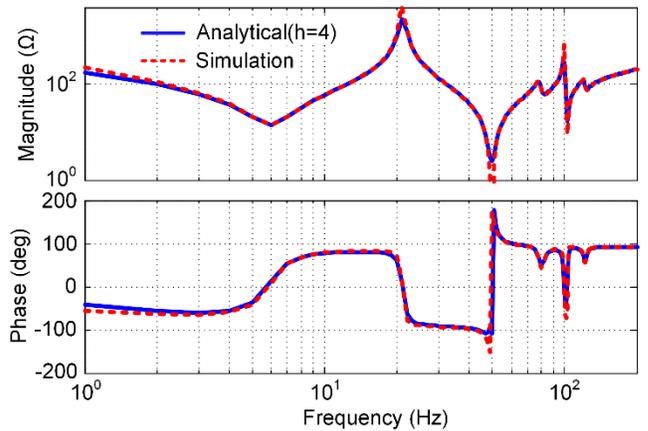

Fig. 7. Comparison between the analytical and measured ac-side impedances of the MMC with ac voltage closed-loop control.

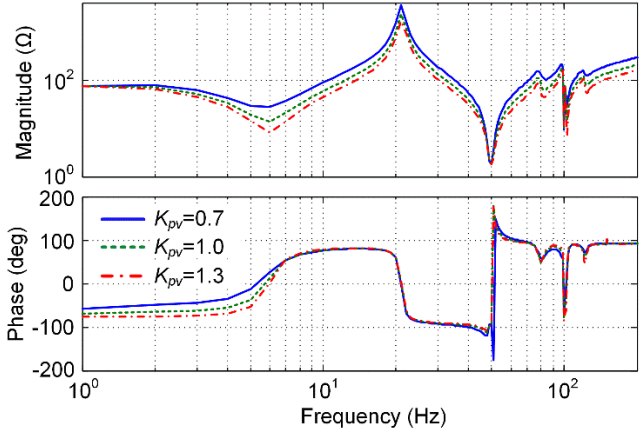

Fig. 8. Impact of the proportional gain of the ac voltage controller on the ac-side impedance of the MMC.

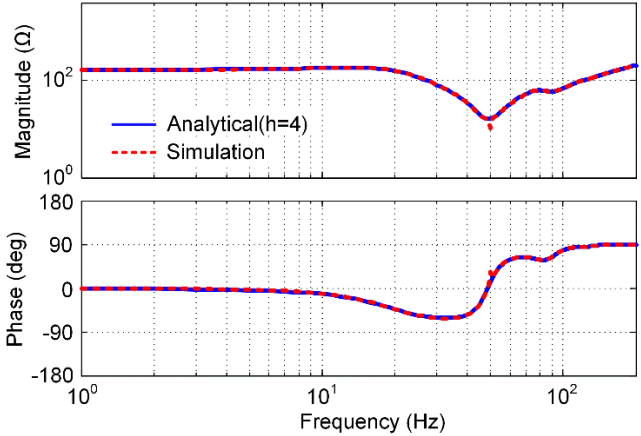

Fig. 9. Comparison between the analytical and measured ac-side impedances of the MMC with circulating current closed-loop control.

Fig. 9 shows the comparison between the analytical and simulation measured ac-side small-signal impedances of the MMC with circulating current closed-loop control (where Ra=20). As can be seen, the analytical impedance has a good agreement with the measured result in the simulation, which validates the analytical impedance model as well. Moreover, it is worth noting that the low-frequency resonance peak is well suppressed by the circulating current controller, which indicates that the circulating current controller can significantly increase the internal damping of the MMC. Thereby, the proportional gain Ra of the circulating current controller can be regarded as an active resistance.

### B. Experimental Verification

To further validate the derived MMC impedance models, the small-signal impedance measurements on a three-phase scale-down MMC experimental setup have also been carried out. The topology of the MMC experimental setup is identical to that in Fig. 1. The California Instruments MODEL RS90 programmable power source is used as the small perturbation voltage injection source which is in series with the three-phase resistance load. The output frequency range of the programmable power source is from 16 Hz to 500 Hz. A dc power source is connected to the dc terminals of the MMC. The schematic diagram of the experimental setup is illustrated in Fig. 10. The main electrical parameters of the experimental setup are as follows: dc voltage $V_{dc}$ = 500V, ac phase voltage amplitude $V_m$ = 200V, SM number per arm $N$ = 12, SM capacitance $C_{SM}$ = 6.6 mF, arm inductance $L$ = 5 mH, load resistance $R_L$ = 10 Ω, and output phase RMS voltage of the programmable power source 7 V.

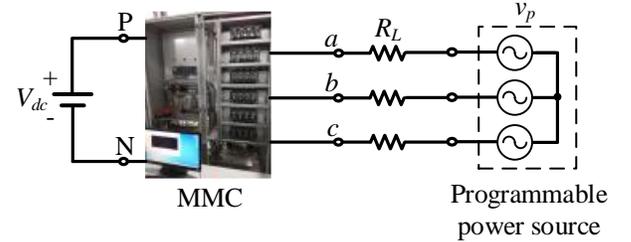

Fig. 10. Schematic diagram of the experimental setup.

It is worth noting that there are relatively significant resistive components in the low-voltage MMC experimental setup, compared with the high-voltage applications. As a result, the internal damping of the low-voltage MMC setup is much stronger than that of the high-voltage one, which means that the instability issues caused by the internal dynamics of MMC are not prominent for the low-voltage MMC setup. This is a key factor to be considered in order to simulate the operational characteristics of the high-voltage MMC setups. Therefore, a virtual resistance compensation (VRC) based strategy is proposed in this work to counteract the actual physical resistance in order to make the operational characteristics of the low-voltage MMC setup appear closer to those of the high-voltage one. As aforementioned the circulating current control strategy essentially increases the arm equivalent series resistance because of the positive proportional gain $R_a$. Whereas if the proportional gain $R_a$ is negative, it can be regarded as a negative resistance that can counteract the arm parasitic resistance. Therefore, the resistance compensation strategy based on the circulating current control with negative $R_a$ in Fig. 4 is used in the course of impedance measurements on the MMC experimental setup.

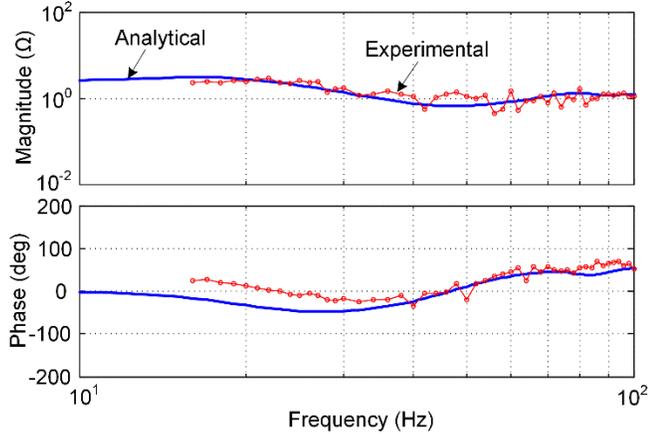

Fig. 11. Experimental measurement impedance with open-loop control and without VRC.

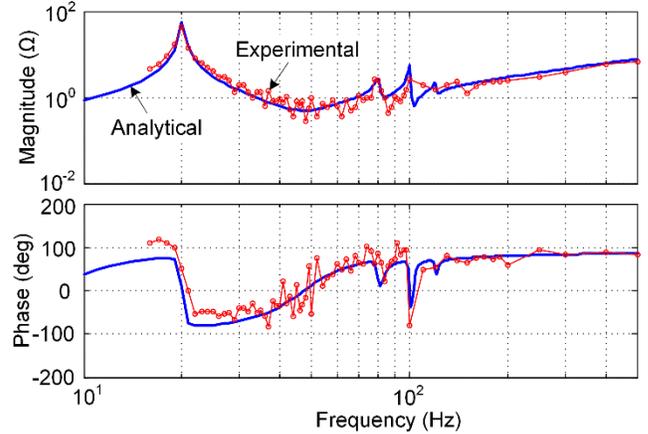

Fig. 12. Experimental measurement impedance with open-loop control and with VRC.

At first, the ac-side small-signal impedance of the MMC with open-loop control was measured in the case of no VRC control, as shown in Fig. 11, where it can be seen that there is a good agreement between the experimental measurement impedance and the analytical model. It is worth noting that there is no low-frequency resonance peak in the MMC impedance, which is due to the relatively large resistive components in the low-voltage MMC experimental setup. Furthermore, the analytical impedance is obtained by setting the arm resistance value $R = 1\ \Omega$, which shows a good matching with the experimental results. It indicates that the arm equivalent series resistance of the MMC experimental setup is approximately equal to $1\ \Omega$. In addition, since the effect of the circulating current control with positive $R_a$ is similar to that of the arm physical resistance, the measurements for the MMC impedance under the circulating current control with positive $R_a$ are no longer given in this paper.

Then, the ac-side small-signal impedances of the MMC with VRC control were measured in the experiments, as shown in Fig. 12 and Fig. 13, where the proportional gain $R_a = -1$. Fig. 12 and Fig. 13 are the comparisons between the analytical and experimentally measured impedances of the MMC with open-loop control and ac voltage closed-loop control, respectively. As can be seen, the agreement between the analytical and experimental measurement impedances is good, and the discrepancies seen may be the results of harmonic interference. It is worth noting that the low-frequency resonance peaks appear in the MMC impedances when the resistance compensation strategy is used, which confirms the effectiveness of the proposed resistance compensation strategy.

## VI. CONCLUSION

This paper presents the small-signal impedance modeling of the MMC. The internal dynamics of the MMC such as harmonic circulating currents and capacitor voltage harmonics, which have great influence on the MMC impedance, must be considered in the impedance modeling. The HSS modeling method is applied to build the harmonic coupling model of the MMC, based on which the small-signal impedance models of the MMC are derived. Three cases, i.e., open-loop control, ac voltage closed-loop control, and circulating current closed-loop control, have been considered in the MMC impedance modeling in order to uncover how the control affects the impedance-frequency characteristics. Finally, both the simulation and experimental measurements verify the effectiveness of the proposed impedance models.

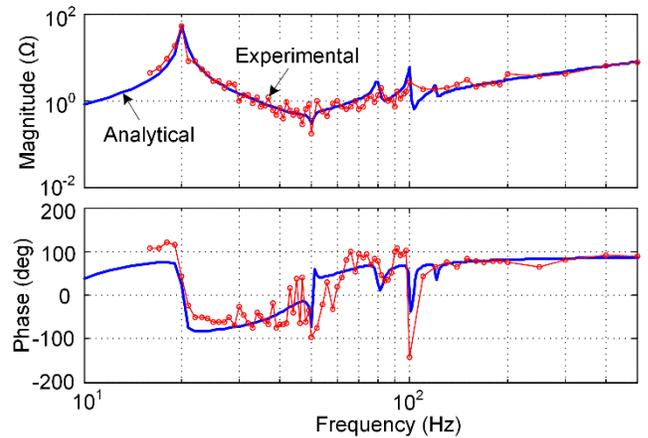

Fig. 13. Experimental measurement impedance with ac voltage closed-loop control and with VRC.